\documentclass[12pt]{iopart}

\usepackage{graphicx,float}

\begin{document}

\title{Many-Body Effects in a Model of Electromagnetically Induced Transparency }

\author{Jose Reslen}

\address{Department of Physics, National University of Singapore, Science Drive 3, Blk 12, Singapore 117543.}%

\date{}

\begin{abstract}
We study the spectral properties of a many-body system
under a regime of electromagnetically induced transparency. 
A semi-classical model is proposed to incorporate the effect
of inter-band interactions on an otherwise single-body scheme.
We use a Hamiltonian with non-Hermitian terms to account for 
the effect of particle decay from excited levels. We 
explore the system response as a result of varying the interaction
parameter. Then we focus on the highly interacting case, also
known as the blockade regime. In this latter case we present
a perturbative development that allows to get the transmission
profile for a wide range of values of the system parameters. We observe
a reduction of transmission when interaction increases and show
how this property is linked to the generation of a strongly correlated
many-body state. We study the characteristics of such a state and
explore the mechanisms giving rise to various interesting features.
\end{abstract}

\maketitle

The study of light--matter interaction 
has experienced a pronounced development over the
last few decades, both experimentally and theoretically. In this
process, progress has been made by advancing 
the foundations laid by pioneering observations, such as electromagnetically 
induced transparency (EIT), stimulated Raman adiabatic 
passage (STIRAP) and coherent population trapping (CPT).
Such techniques can be understood in terms of
single particle models, making it possible to
obtain much insight into the physical background leading
to the realisation of the mentioned phenomena. Similarly, due to 
the increasing level of control possible in low-temperature physics, 
the transition toward more collectively driven scenarios is
starting to attract general attention. 

\begin{figure}[H]
\begin{center}
\includegraphics[width=0.5\textwidth,angle=-0]{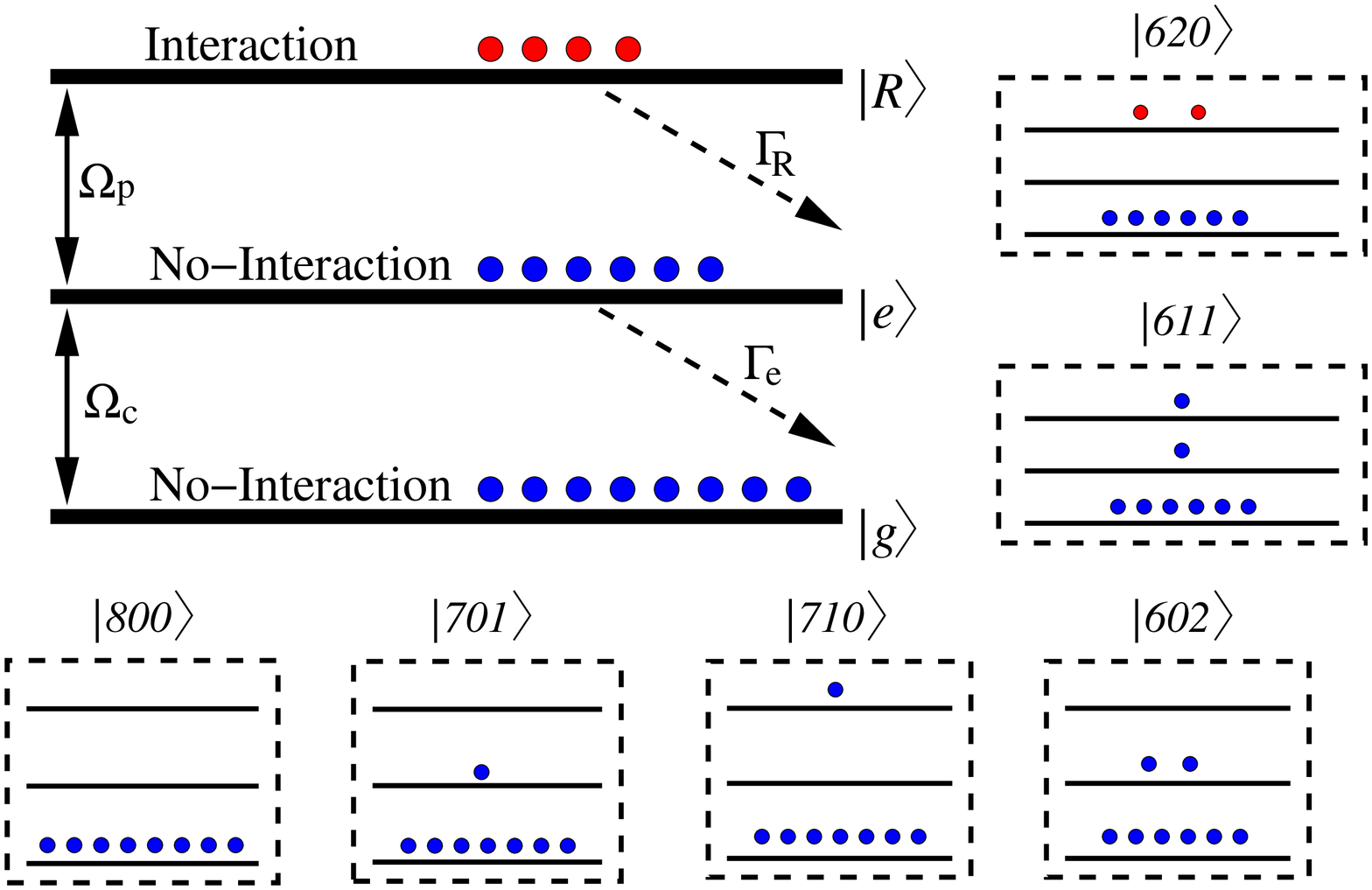}\\
\caption{Diagram of EIT including several particles. The dynamics is 
determined by the coupling and probe lasers which drive particles 
between the ground and excited bands and the excited and 
interaction (Rydberg) bands respectively. In addition, the first
excitation levels of a system with 8 bosons are shown.} 
\label{fig1}
\end{center}
\end{figure}

One way of understanding the mechanism underlying EIT is to
think in terms of cancellation of excitation pathways with 
opposite phases \cite{Marangos}. As the process involving the
absorption of one photon is on-resonance, a different transition
enhancing photon emission arises with equal intensity and opposite phase. The
result can then be seen as pure quantum interference. An
alternative way is to view the optical field as interacting
with just one of two possible coupling modes \cite{Renzoni}. 
The resonant mode is related to the so-called dark state, which
displays a decay time much lower than other possible configurations,
which allows for the whole system to be pumped into this dark state
by the combined action of probe and coupling lasers.
These views are useful when many-body effects
are taken into consideration, but signatures of collective
behaviour slowly degrade the single particle response \cite{Plotz},
 and a more detailed analysis becomes necessary. This response
results as a consequence of the interaction among particles 
and is an instance in which photons communicate using matter
as an intermediary \cite{Alexey}. In this direction, an interesting problem 
is the study of the many-body physics behind the interference 
mechanism governing EIT. An example of this is the realisation
of clean EIT-profiles on ensembles of interacting atoms \cite{Pritchard,Schempp}.

Atoms with very high principal quantum number---also known
as Rydberg atoms--display important properties that make
them useful in several applications. In particular,
the interaction among Rydberg atoms gives rise to a
blockade mechanism originated by dipole--dipole interactions or
van der Waals forces. The blockade process can be exploited in diverse applications, 
such as quantum gates proposals \cite{efrain,inbal,Muller,Wu,Isenhower}, cavity 
QED constructs \cite{Guerlin,yuan}, the generation of entanglement \cite{Wilk,Gillet,kwek}, 
adiabatic passage \cite{Moller,Beterov} and numerous quantum information techniques \cite{Saffman}.

For the specific case of EIT, inter-atomic interactions
produce a highly robust state with a collective character.
As discussed in reference \cite{Pritchard}, the quantum superposition
can be seen as involving states where due to dipole--dipole interactions
only a single atom can be
excited to the interacting band and therefore only one atom
is involved in EIT. This results in a reduction of transparency
as the possibilities of interference of excitation pathways 
are diminished. Such reduction shows neither resonance-shift nor
line-width broadening. Some of the experimental results cannot
be explained via mean-field approximations and 
full quantum models must be employed to reproduce the observed features \cite{Schempp}.
In reference \cite{Otter} it was shown that two-photon correlations lead
to an enhanced attenuation of the probe beam for strong intensities
and in this way a detailed description of the experiment
in reference \cite{Pritchard} is achieved. When emphasis is made on the propagation
of the probe laser through the medium as in \cite{Henkel}, it is
shown that the blockade mechanism gives rise to a highly non-local
response in addition to non-linearities.

Many-body effects in an ensemble of interacting atoms can be studied
using the master equation formalism \cite{Ates}. In this case the  number of 
coefficients necessary to describe the density matrix is proportional to the square of the 
total number of elements in the Hilbert space. Instead, we probe the 
advantages of introducing decay factors as imaginary elements in 
the Hamiltonian, so that the whole analysis can be carried in terms
of state functions, providing insight into the development of the many-body 
scenario as well as the statistical effects that arise due to the
bosonic nature of the particles. Non-Hermitian Hamiltonians (NHH) 
have proved useful in several studies, {\it e.g.}, single particle 
models of EIT \cite{Marangos}, STIRAP in the presence of degenerate 
product states \cite{Gong1,Gong2} and the dynamics of Bose--Hubbard 
dimers \cite{Graefe} among many others. This approach is
reasonable since we want to explore the case of high
number of particles where standard approximations have displayed
mixed results \cite{Brazil}. Below we show that the proposed 
methodology produce consistent results
and is especially suitable to develop a perturbative approach which
is valid over a wide range of parameters. The insight acquired in this
way is less accessible using a fully numerical approach since several tunable
parameters must be considered.

In our proposal we assume that probe and coupling lasers induce particle exchange among
three energy levels as in a ladder EIT scheme \cite{Pritchard,Moller} 
(figure \ref{fig1}) and that two-body interactions take place only in the interaction band.
Such two-body exchange is mediated by a constant parameter $U$. 
In general, the interaction depends on the inter-atomic distance
and the Rydberg principal quantum number \cite{Pritchard,Saffman}.
Using a semi-classical approach and incorporating the rotating wave 
approximation in addition to non-Hermitian decay-terms the NHH reads,
\begin{eqnarray}
& \hat{H} = - \Omega_c e^{-i \omega_c t} \hat{a}_R^{\dagger} \hat{a}_e  - \Omega_p e^{-i \omega_p t} \hat{a}_g^{\dagger} \hat{a}_e + H.c. \nonumber & \\
& + \frac{U}{2}\hat{a}_R^{\dagger}\hat{a}_R(\hat{a}_R^{\dagger}\hat{a}_R-1) - i \Gamma_R \hat{a}_R^{\dagger}\hat{a}_R - i \Gamma_e \hat{a}_e^{\dagger}\hat{a}_e \nonumber & \\
& + E_g \hat{a}_g^{\dagger} \hat{a}_g + E_R \hat{a}_R^{\dagger} \hat{a}_R + E_e \hat{a}_e^{\dagger} \hat{a}_e.
\label{eq:1}
\end{eqnarray}

Creation and annihilation operators introduce boson-like statistics
through the commutation relations $[\hat{a}_L,\hat{a}_L^{\dagger}]=1$ for $L=g,e,R$.
$\Omega_p$ and $\Omega_c$ are the probe- and coupling-Rabi frequencies 
respectively. Similarly, $\omega_p$ and $\omega_c$ are the laser frequencies
while $E_g$, $E_R$ and $E_e$ represent the energies of the corresponding 
bands. 
The intensity of particle decay from the interaction- and excited-bands
is controlled via $\Gamma_R$ and $\Gamma_e$ respectively. Both constants are 
positive definite. This approach is valid in the limit 
of weak coupling between the ground and excited bands \cite{Marangos}. 
In our analysis we set $\hbar = 1$ and measure all the parameters
in recoil energies $E_r=\hbar^2 K^2/(2m)$. The energy structure of
the system correspond to three-level EIT-picture where 
$E_R>E_e>E_g$ (figure \ref{fig1}).
Although not explicit in (\ref{eq:1}), the total number of particles
is $M$, which accounts for the number of atoms in a blockade sphere. 
The proposed scheme can be realised by projecting counter-propagating
lasers onto a cloud of ultra-cold atoms---for example, ${}^{87}$Rb. These lasers
provide the probe- and coupling-frequencies in our proposal. The number
of atoms can be controlled by pumping atoms into energy levels that
do not couple to the probe- or coupling-lasers. The energy spectrum
of the gas contains bands that can stand for the ground and excited
bands of figure \ref{fig1}. The gas also contains highly excited
atoms which display interaction intensities much larger than 
atoms in the ground or excited levels, so that it is valid to
neglect interaction effects on these bands. The intensity of the interaction
can be tuned using Feshbach resonance \cite{Wieman}. Actual experiments 
implementing this approach have been reported in several works,
as for instance in references \cite{Pritchard,Schempp}. 
As a result of the inclusion of non-Hermitian terms in equation (\ref{eq:1}), the wave function norm is no longer preserved
and therefore the quantum state must be normalised
in anticipation to any explicit calculation. After an appropriate 
transformation we get the following dressed NHH in the
interaction picture,
 \begin{eqnarray}
& \hat{H}_{D} = - \Omega_c (\hat{a}_R^{\dagger} \hat{a}_e + \hat{a}_e^{\dagger} \hat{a}_R) - \Omega_p (\hat{a}_g^{\dagger} \hat{a}_e + \hat{a}_e^{\dagger} \hat{a}_g)  \nonumber & \\
& + \frac{U}{2}\hat{a}_R^{\dagger}\hat{a}_R(\hat{a}_R^{\dagger}\hat{a}_R-1) - i \Gamma_R \hat{a}_R^{\dagger}\hat{a}_R - i \Gamma_e \hat{a}_e^{\dagger}\hat{a}_e  \nonumber & \\
& + \delta \hat{a}_g^{\dagger} \hat{a}_g + (E_R - \omega_c)  \hat{a}_R^{\dagger} \hat{a}_R + E_e \hat{a}_e^{\dagger} \hat{a}_e,
\label{eq:2}
\end{eqnarray}
where $\delta = E_g - \omega_p$. For simplicity we have assumed single photon
resonance and without loss of generality we set 
$E_R - \omega_c = E_e = 0$. Here we are mainly concerned with 
the magnitude of the coupling between the atoms and the probe laser, i.e.,
the atomic susceptibility. Hence we focus on the mean value,
\begin{equation}
\chi_M^{(n)} = \frac{\langle  \left( \hat{a}_g^{\dagger} \hat{a}_e \right)^n  \rangle}{M^n}.
\label{eq:18}
\end{equation}

\begin{figure}
\begin{center}
\includegraphics[width=0.5\textwidth,angle=-90]{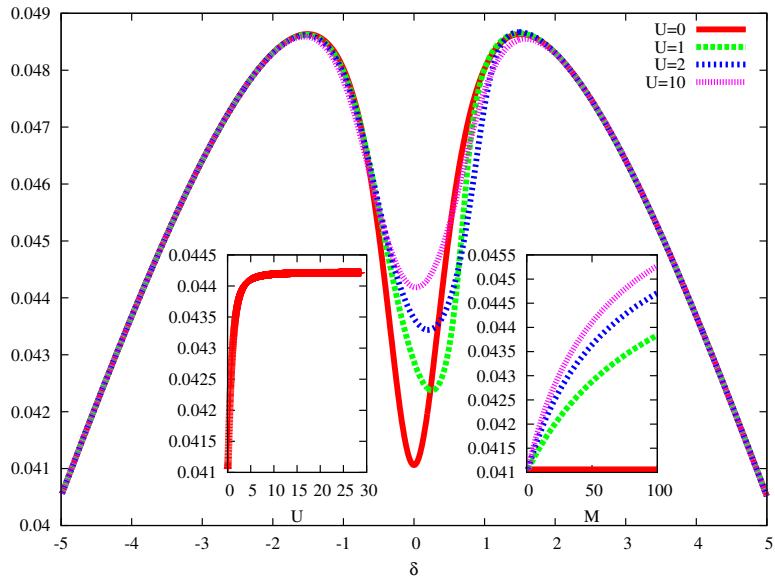}\\
\caption{$\chi_M^{(1)}$ for $M=50$, $\Gamma_e=10$, $\Gamma_R=0.5$, $\Omega_c=1$, 
$\Omega_p=0.5$. Inset: Same parameters and $\delta=0$. Left: $\chi_M^{(1)}$ 
against $U$. Right:$\chi_M^{(1)}$ against $M$. The large dip in the main figure 
features a decrease of absorption of the probe laser at zero detuning. Increasing
the interaction provokes a reduction of transmission. As shown in the insets, 
this effect can be controlled by tuning either the intensity of the interaction 
or the number of particles.} 
\label{fig:2}
\end{center}
\end{figure}

Imaginary and real parts of $\chi_M^{n}$ account for different orders of
absorption and refraction respectively. Figure \ref{fig:2} presents 
the behaviour of $\chi_M^{(1)}$ as various parameters are tuned.
In all cases the state of the system is given by the right eigenstate 
of equation (\ref{eq:2}) corresponding to the eigenvalue with the highest 
imaginary part (the less decaying state). The NHH (\ref{eq:2}) conserves 
the total number of particles so that the full dimension of the Hilbert 
space is $(M+2)(M+1)/2$.
The pattern shown in figure \ref{fig:2} is proportional to the
transmission profile of the probe laser. It indicates a characteristic
window of transparency that results from the
combined action of the incident lasers. In the
absence of interaction the dip in absorption can be ascribed to
interference of pathways with opposite phases, {\it i.e.}, the process
by which one atom goes from $|g\rangle$ to $|e\rangle$ is 
cancelled out by the process by which the atom goes from $|e\rangle$
to $|R\rangle$ and then all the way back from $|R\rangle$ to $|e\rangle$ 
to $|g\rangle$ again. As can be seen, the case $U=0$, 
which is equivalent to the single particle case, displays
the maximum interference. As the interaction is gradually turned on, the
two-photon resonance is shifted to the right due to the energy increase of 
the interaction band produced by the repulsion among particles. 
$\chi_M$ continues to grow until it reaches a 
saturation value, but without completely suppressing transparency. 
$\chi_M$ also grows with $M$, but no saturation is visible
over values of $M$ less than 100. In general, we
can see that the insertion of particles causes a reduction of
quantum interference. Such an effect can be enhanced either by
increasing $U$ or $M$. However, in every case the physical
response is different. While changing $U$ affects the intensity
of two-body interactions, adding particles produces a (sometimes steep)
rearrangement of the quantum state.

From the Heisenberg equations we can extract the expression,
\begin{eqnarray}
\frac{d\hat{\alpha}_g^{\dagger}}{dt} = i\Omega_p \hat{\alpha}_e^{\dagger} - i \delta \hat{\alpha}_g^{\dagger},
\label{eq:3}
\end{eqnarray}
where $\hat{\alpha}_n^{\dagger} = e^{-it\hat{H}_D} \hat{a}_n^{\dagger} e^{it\hat{H}_D}$ for $n=g,R,e$. Since
$\Omega_p << \Omega_c$ we can assume that $\hat{\alpha}_g^{\dagger}$ does not
greatly influence the evolution of $\hat{\alpha}_R^{\dagger}$ and $\hat{\alpha}_e^{\dagger}$, so that 
$\hat{\alpha}_e^{\dagger}$ can be treated as a function of time. Thus it follows  that,
\begin{equation}
\hat{\alpha}_g^\dagger(t) = i \Omega_p e^{-i \delta t} \int_0^t e^{i \delta T} \hat{\alpha}_e^{\dagger}(T) dT.
\label{eq:4}
\end{equation}

Under such an assumption, in the blockade regime equation (\ref{eq:2}) can 
be reduced to a Jaynes--Cummings-like Hamiltonian on the interaction- and excited-bands, 
\begin{equation}
\hat{H}_{Re} = -\Omega_c (\hat{\sigma}^+ \hat{a}_e + \hat{\sigma}^- \hat{a}_e^{\dagger} ) - i \Gamma_R \left( \frac{\hat{\sigma}_z+1}{2} \right) -i \Gamma_e \hat{a}_e^{\dagger} \hat{a}_e,
\label{eq:5}
\end{equation}

where we have employed Pauli matrices to account for the dynamics
of the interaction band. The eigensystem of equation (\ref{eq:5}) is given by \cite{comment2},

\begin{eqnarray}
&E=0, \hspace{0.5cm} |0,0\rangle, \hspace{0.5cm} \langle 0,0|, & \\
&E_n^{\pm}=-i(\Gamma_e n + \Gamma_R + \Omega_c \sqrt{n+1} e^{\pm i\theta_n}), \label{eq:6}& \\
&|E_n^{\pm} \rangle = (|1,n\rangle + i e^{\pm i \theta_n} |0,n+1\rangle)/Z_n^{\pm}  , \label{eq:7} &\\
& \langle E_n^{\pm} | = (\langle 1,n | + i e^{\pm i \theta_n} \langle 0,n+1 |)/Z_n^{\pm}, \label{eq:75} &
\end{eqnarray}
where,
\begin{eqnarray}
& Z_n^{\pm} = \sqrt{1-e^{\pm 2i\theta_n}},   \label{eq:78} & \\
&\cos \theta_n = \frac{\Gamma_e-\Gamma_R}{2\Omega_c \sqrt{n+1}},\hspace{0.2cm} n = 0,1,\dots,M-1.  \label{eq:8}&
\end{eqnarray}

The first and second integers in a ket (bra) correspond to the
number of particles in the interaction (excited) band.
The angle $\theta_n$ is complex as well as the eigenvalues
of the NHH. In the form in which they appear above, the eigenvectors satisfy 
$\langle E_k^s | E_{k'}^{s'} \rangle = \delta_k^{k'} \delta_s^{s'}$. 
This eigensystem can be used to write,

\begin{equation}
\hat{H}_{Re} = \sum_{n=0}^{M-1} E_n^{\pm} |E_n^{\pm} \rangle \langle E_n^{\pm}|.
\label{eq:134}
\end{equation}

The spectrum of the NHH is well defined except when $\cos \theta_n = 1$, 
in which case the eigenvectors (\ref{eq:7}) and (\ref{eq:75}) become 
indeterminate but one can still recover the NHH as the limit
of equation (\ref{eq:134}). These singularities therefore produce removable
discontinuities leading to no divergence in the mean values of the 
system's observables. Now we project on the subspace associated to the NHH,

\begin{equation}
\hat{a}_e^{\dagger} = \hat{a}_e^{\dagger} \left \{ \sum_{n=0}^{M-1} |E_n^{\pm} \rangle \langle E_n^{\pm}|  + |0,0\rangle \langle 0,0 | \right \}.
\end{equation}

This procedure is facilitated by introducing the coefficients
validating the following identities:

\begin{eqnarray}
& \hat{a}_e^{\dagger}  |0,0 \rangle = c_{0}^+ |E_{0}^+ \rangle + c_{0}^- |E_{0}^- \rangle,& \\
& \hat{a}_e^{\dagger}  |E_n^+ \rangle = c_{n+1}^+ |E_{n+1}^+ \rangle + c_{n+1}^- |E_{n+1}^- \rangle, &  \\
& \hat{a}_e^{\dagger}  |E_n^- \rangle = d_{n+1}^+ |E_{n+1}^+ \rangle + d_{n+1}^- |E_{n+1}^- \rangle. &
\end{eqnarray}

Once $\hat{a}_e^{\dagger}$ is written in terms of the eigenvectors of the NHH we can calculate 
$\hat{\alpha}_e^{\dagger}$ and replace it in equation (\ref{eq:4}) where we can now 
carry on the integration in terms of time. The result can be written in the form,

\begin{equation}
\hat{\alpha}_g^{\dagger}(t) =  \hat{a}_g^{\dagger} + \Omega_p \left ( \hat{\kappa}(0) - e^{-i \delta t} \hat{\kappa}(t) \right ),
\label{eq:201}
\end{equation}

where,

\begin{eqnarray}
 \hat{\kappa}(t) =  \frac{c_{0}^+ e^{-i t (E_{0}^+ - \delta)} | E_{0}^+ \rangle
\langle 0,0 | }{E_{0}^+ - \delta} + \frac{c_{0}^- e^{-i t (E_{0}^- - \delta)} |
E_{0}^- \rangle \langle 0,0 | }{E_{0}^- - \delta} +   && \nonumber \\
\sum_{n=0}^{M-2} \left (
\frac{c_{n+1}^+ e^{-i t (E_{n+1}^+ - E_n^+ - \delta)} | E_{n+1}^+ \rangle
\langle E_n^+ | }{E_{n+1}^+ - E_n^+ - \delta} +    \right .   
 \frac{c_{n+1}^- e^{-i t (E_{n+1}^- - E_n^+ - \delta)} | E_{n+1}^- \rangle \langle E_n^+ | }{E_{n+1}^- - E_n^+ - \delta} + && \nonumber \\
\frac{d_{n+1}^+ e^{-i t (E_{n+1}^+ - E_n^- - \delta)} | E_{n+1}^+ \rangle \langle E_n^- | }{E_{n+1}^+ - E_n^- - \delta} +    
\left . \frac{d_{n+1}^- e^{-i t (E_{n+1}^- - E_n^- - \delta)} | E_{n+1}^- \rangle \langle E_n^- | }{E_{n+1}^- - E_n^- - \delta} \right ).  && 
\label{eq:202}
\end{eqnarray}

Let us assume that time-dependent terms decay over time. The range of validity
of this assumption will be addressed further below. Since at $t=0$ all the particles
remain in the ground band the stationary state can be obtained from $\hat{\alpha}_g^{\dagger}$
in accordance to the Heisenberg picture,

\begin{equation}
\left(\hat{\alpha}_g^{\dagger}(\infty)\right)^M |0\rangle = \sum_{k=0}^M
\left(
\begin{array}{c}
M \\
k
\end{array}
\right)
\left(\hat{a}_g^\dagger \right)^{M-k} (\Omega_p \hat{\kappa}(0))^k |0\rangle.
\label{eq:10}
\end{equation}

Operator $\hat{\kappa}(0)$ couples the representations of $\hat{H}_{Re}$ in such a way that,

\begin{equation}
\hat{\kappa}(0)^k |0\rangle = v_k Z_{k-1}^+ |E_{k-1}^+ \rangle + w_k Z_{k-1}^- |E_{k-1}^- \rangle, k=1,\dots,M.
\label{eq:11}
\end{equation}

By replacing (\ref{eq:11}) in (\ref{eq:10}) one obtains,

\begin{eqnarray}
&& |M00\rangle + \sum_{k=1}^M \sqrt{
\left(
\begin{array}{c}
M \\
k
\end{array}
\right)
} \Omega_p^k \left ( (v_k+w_k)|M-k,1,k-1\rangle \right . \nonumber \\
&& \left .  + i(v_k e^{i \theta_{k-1}} + w_k e^{-i \theta_{k-1}}) |M-k,0,k\rangle \right ).
\label{eq:12}
\end{eqnarray}

The set of coefficients $v_k,w_k$ are connected by the recursion expression,

\begin{equation}
\left(
\begin{array}{c}
v_{k} \\
w_{k}
\end{array}
\right)
=
\hat{B}_k
\left(
\begin{array}{c}
v_{k-1} \\
w_{k-1}
\end{array}
\right),\hspace{0.2cm} k=2,\dots,M.
\label{eq:13}
\end{equation}

From an induction argument $\hat{B}_k$ is found to satisfy,
\begin{eqnarray}
&&  \hat{B}_k 2i\sqrt{k} \sin \theta_{k-1}= \label{eq:25}\\
&& \left(
\begin{array}{cc}
\frac{-\sqrt{k-1}e^{-i\theta_{k-1}} + \sqrt{k}e^{i\theta_{k-2}} }{(E_{k-1}^+ - E_{k-2}^+ - \delta)} & \frac{ -\sqrt{k-1}e^{-i \theta_{k-1}} +\sqrt{k}e^{-i\theta_{k-2}}}{(E_{k-1}^+ - E_{k-2}^- - \delta)} \\ 
 \frac{\sqrt{k-1}e^{i \theta_{k-1}}-\sqrt{k}e^{i\theta_{k-2}} }{(E_{k-1}^- - E_{k-2}^+ - \delta)} & \frac{\sqrt{k-1}e^{i\theta_{k-1}}-\sqrt{k}e^{-i\theta_{k-2}}}{(E_{k-1}^- - E_{k-2}^- - \delta)}   
\end{array}
\right), \nonumber
\end{eqnarray}
and the first elements are given by,
\begin{equation}
v_1 = \frac{-1}{2 (E_0^+-\delta) \sin \theta_0}, \hspace{0.3cm} w_1 = \frac{1}{2(E_0^--\delta)\sin \theta_0}. 
\label{eq:15}
\end{equation}

Equations (\ref{eq:15}), (\ref{eq:25}), (\ref{eq:13}) and (\ref{eq:12}) can be used to numerically
generate the stationary state for any set of parameters excluding singularities.
Such a condition can be improved by introducing new discrete variables,
\begin{equation}
p_k = v_k + w_k, \hspace{0.5cm} q_k = \sin \theta_{k-1} (v_k - w_k).
\label{alex}
\end{equation}

\begin{figure}
\begin{center}
\includegraphics[width=0.5\textwidth,angle=-90]{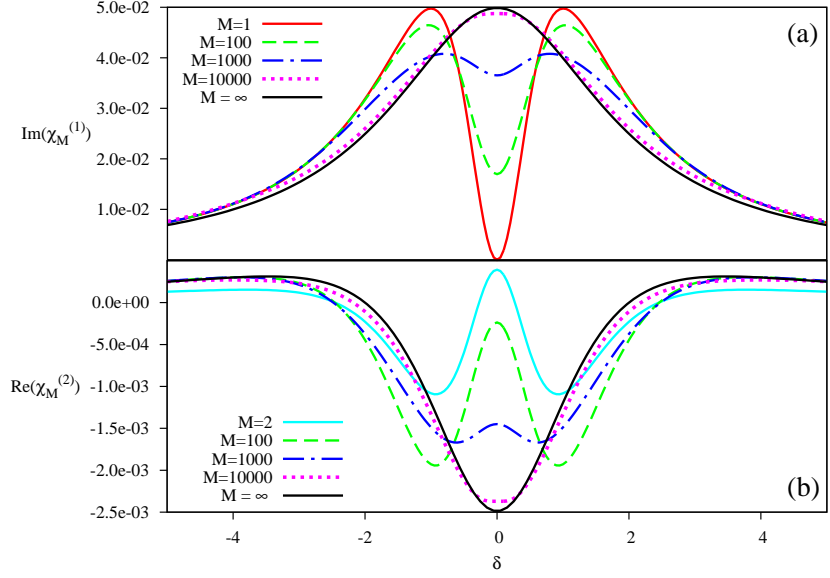}
\caption{First order absorption (a) and second order refraction (b) profile for different $M$. The cases $M=1$ and $M=\infty$ correspond to equations (\ref{eq:19}) and (\ref{eq:20}) respectively. Other cases shown are generated following the procedure sketched in the text. The parameters involved in the computations are $\Gamma_e=2$, $\Gamma_R=0$, $\Omega_p=0.1$ and $\Omega_c=1$.} 
\label{fig:4}
\end{center}
\end{figure}

These variables obey a recursion relation analogous to equation (\ref{eq:13})
with $v_k$ and $w_k$ replaced by $p_k$ and $q_k$ respectively and 
$\hat{B}_k$ replaced by,
\begin{eqnarray}
&& \hat{A}_k = \frac{1}{\sqrt{k}F_k} \left(
\begin{array}{cc}
 R_k \omega + S_k T_k \sin^2 \theta_{k-2} & R_k T_k + S_k \omega   \\ 
 r_k \omega + s_k T_k \sin^2 \theta_{k-2} &  r_k T_k + s_k \omega  
\end{array}
\right), \nonumber
\end{eqnarray}
in such a way that,

\begin{eqnarray}
& F_k = \omega^2 - T_k^2 \sin^2 \theta_{k-2}, & \\
& R_k = i \Omega_c \sqrt{\frac{k}{k-1}} \cos \theta_{k-1} - \eta \sqrt{k-1}, & \\
& S_k = -\Omega_c(2k -1), & \\
& r_k = \frac{i\eta \cos \theta_{k-1}}{\sqrt{k-1}} - \Omega_c \sqrt{k(k-1)} \left( \sin^2 \theta_{k-1} + \sin^2 \theta_{k-2} \right), & \\
& s_k = i \Omega_c \cos \theta_{k-1} - \eta \sqrt{k}, & \\
& T_k = -2 \eta \Omega_c \sqrt{k-1}, & \\
& \omega = \eta^2 - \Omega_c^2, & \\ 
& \eta = i\Gamma_e + \delta. &
\label{eq:324}
\end{eqnarray}

These expressions along with the initial coefficients,

\begin{eqnarray}
p_1 = \frac{\Omega_c}{(i\Gamma_R + \delta)(i\Gamma_e + \delta) - \Omega_c^2}, \\
q_1 = \frac{(i(\Gamma_e + \Gamma_R) + 2\delta)/2}{(i\Gamma_R + \delta)(i\Gamma_e + \delta) - \Omega_c^2},
\label{423}
\end{eqnarray}

\begin{figure}
\begin{center}
\includegraphics[width=0.5\textwidth,angle=-90]{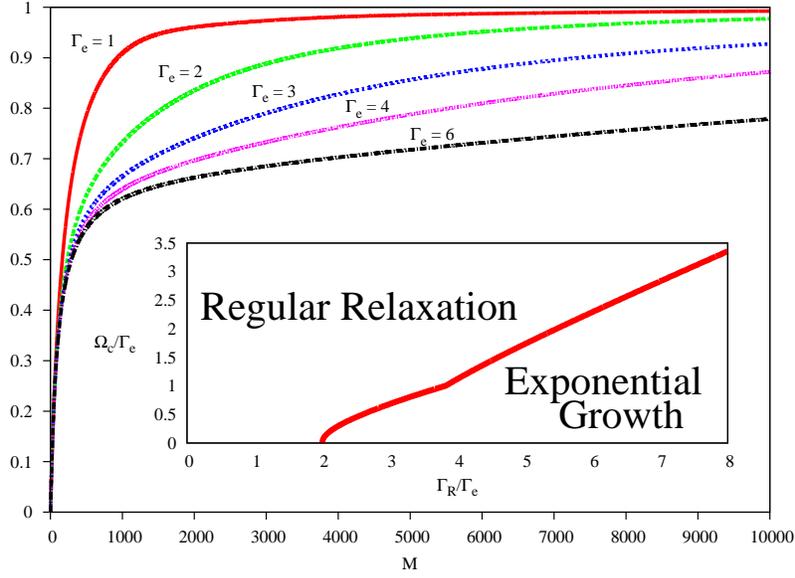}\\
\caption{$\frac{Im \left ( \chi_M^{(1)}  \right  )}{Im \left( \chi_{\infty}^{(1)} \right)}$. In every case 
$\delta=0$, $\Gamma_R=0$, $\Omega_c=1$ and $\Omega_p=0.1$. 
The absorption curves display non-linear signatures.
Saturation takes place according to the interplay between decay and coupling. 
Inset: Relaxation map for $M>1$. The case $M=1$ shows no signatures of 
exponential growth.} 
\label{fig:3}
\end{center}
\end{figure}

can be integrated into a programming routine that recursively calculates 
the state coefficients and then $\chi_M^{(n)}$. We note that when
$|\Gamma_e-\Gamma_R|<<2 \Omega_c k$ the 
recursion matrix can be approximated as follows,

\begin{equation}
\hat{A}_k \approx
\left(
\begin{array}{cc}
\frac{-1}{i\Gamma_e+\delta} & 0 \\
0 & \frac{-1}{i\Gamma_e+\delta}
\end{array}
\right).
\label{eq:16}
\end{equation}

Employing this $\hat{A}_k$ one finds a stationary state displaying 
the following susceptibility \cite{comment3},

\begin{equation}
\chi_{\infty}^{(1)} = \frac{-\Omega_p}{(i\Gamma_e+\delta)}, \hspace{0.5cm} \chi_{\infty}^{(2)} = \left ( \chi_{\infty}^{(1)} \right)^2 .
\label{eq:19}
\end{equation}

Similarly, from a direct calculation the single particle susceptibility is found to be,

\begin{equation}
\chi_{1}^{(1)} = \frac{-\Omega_p (i\Gamma_R + \delta)}{(i\Gamma_R + \delta)(i\Gamma_e + \delta) - \Omega_c^2}.
\label{eq:20}
\end{equation}

As can be seen in figure \ref{fig:4}, equations (\ref{eq:19}) and (\ref{eq:20}) 
are extreme cases of the numerical results obtained using the matrix
$\hat{A}_k$ recursively. Here we have chosen the case $\Gamma_R=0$ because
it is most close to the actual experimental case where the decay rate of
the Rydberg state is usually negligible. The inclusion of more particles
in the system induces a rise of $Im\left( \chi_M^{(1)} \right)$ at zero 
detuning. In the same instance, $Re\left( \chi_M^{(2)} \right)$ undergoes 
a change of sign as it dips from its peaking value.
From figure \ref{fig:3} we can see that the absorption profile at
$\delta=0$ features non-linear
behaviour over a wide range of values of $M$ and then asymptotically converges
toward the estimation given by equation (\ref{eq:19}). As we increase $\Gamma_e$, 
the growth speed decreases and the curves display stronger inflection. 
In every case the absorption profile is characterised by a steep absorption 
growth. Such a non-linear response is linked to a cooperative many-body 
state where the single particle outcome is no longer dominant. 
As $M$ goes up further, the absorption value starts to saturate,
suggesting that less particles are being integrated in the
interaction process.

In the single particle case, maximal transmission is achieved
as a result of almost perfect quantum interference between
excitation pathways. In this case the only coupling mode 
involved in EIT in equations (\ref{eq:6}-\ref{eq:8}) is $n=0$. 
Gradual increase of $\Omega_p$ allows more particles into 
the excited band and therefore more coupling modes 
participate in the EIT process. One important characteristic
of EIT is that the reduction of absorption at zero detuning
is accompanied by a peak in the refraction coefficient\cite{Marangos}. 
This allows for powerful applications such as slow light or light storage. 
The mechanism behind the latter procedures is based on the fact
that at $\delta=0$ absorption can be made low at the same time
that the coefficient that determines the velocity of light in
the medium peaks. This important property can be observed without
much effort using our approach. Figure \ref{fig:4}a shows a dip 
in $Im\left( \chi_M^{(1)} \right)$ while figure \ref{fig:4}b shows and a peak in 
$Re\left( \chi_M^{(2)} \right)$ for the same parameters. 
Also at single-photon resonance it can be shown that independently of $M$,

\begin{equation} 
Re\left( \chi_M^{(1)} \right)_{\delta=0} = Im\left( \chi_M^{(2)} \right)_{\delta=0} = 0,
\label{pi}
\end{equation}

Notice that the peaking of refraction can only be seen for $M>1$
because $\chi_1^{(2)}=0$. This means that the system's refraction 
at $\delta=0$ depends only on two-or-more-body transitions, 
where several atoms are excited (or decay) simultaneously. 
In a sense, one can think of
the process taking place at zero detuning as one in which
single-body transitions play a rather marginal role
and instead the leading response is mediated by 
higher order transitions. Likewise, coupling modes involving more than
one particle display less interference, as many particle
pathways between the ground- and excited-bands can only 
interfere partially with their blockade counterparts. 

As even more atoms are integrated in the model,
the possibilities of light being absorbed via particle
excitation become higher and the characteristic profile of EIT
finally fades away. The enhancement of absorption due to 
many-body effects is especially notorious in the range $M<1000$. 
Since the peak of refraction at $M=2$ is positive one can find a
value of $M$ ($M=65$ for the values of figure \ref{fig:4}) for 
which refraction is almost zero at zero-detuning, while absorption is still low. In this case
the system is almost unresponsive to the incoming
radiation. 

In order to check for the consistency of our approach we stress that
our central assumption is the cancellation of $\hat{\kappa}(t)$ in
equation (\ref{eq:202}) as $t\rightarrow\infty$. This is indeed the case 
as long as all the imaginary parts of the arguments of the exponentials in
equation (\ref{eq:202}) turn out to be negative. Only the following arguments
could become greater than zero,

\begin{equation}
Im(E_{n+1}^+ - E_n^-) = \frac{1}{2} \left(  D(n) + D(n+1)  \right) -\Gamma_e, \label{tango4} 
\end{equation}

and,

\begin{equation}
Im(E_{n+1}^- - E_n^-) =  \frac{1}{2} \left(  D(n) - D(n+1)    \right)-\Gamma_e,  \label{tango6}
\end{equation}

where,
 
\begin{equation}
D(n)=Re \left( \sqrt{ (\Gamma_e - \Gamma_R)^2 - 4 \Omega_c^2 (n+1)} \right),
\label{liz}
\end{equation}

for $n=2,3,\cdots,\infty$.

If either (\ref{tango4}) or (\ref{tango6}) become positive then
operator $\hat{\kappa}(t)$ will display exponential growth and will
dominate the stationary state. While it seems operationally
possible to obtain such a state, this feature is less consistent with
our initial consideration in which $\Omega_p$ is a perturbative
parameter and hence most particles remain in the ground band.
Due to the form of $D(n)$, the maximum value of (\ref{tango4}) and 
(\ref{tango6}) takes place at $n=2$. Therefore, if either (\ref{tango4}) or 
(\ref{tango6}) are positive for any $n>2$ then they are positive
for $n=2$ as well. Hence $n=2$ is the only relevant mode in a relaxation
analysis. We have depicted in the inset of 
figure \ref{fig:3} the relaxation map that results following the
arguments discussed above. It is worth pointing out that any arbitrary 
set of realistic parameters in which
$\Gamma_e>\Gamma_R$ fall well inside the regular relaxation zone.
It also becomes apparent that in the many-body case the mere
existence of a dark state does not guarantee the state convergence
toward such a state.

We have investigated the reduction of transparency in an EIT
set-up as a result of the collective character developed by
many-body matter. Results corresponding to a semi-classical 
model were obtained from numerical diagonalisation as well as
from a perturbative approach. In the latter case we presented 
a semi-analytical procedure
that can be used to find the stationary state of the
system in the limit of small intensity of the probe laser.
We have also studied the range of validity of our method and
established explicit conditions for regular relaxation.
The procedure itself shows interesting issues and is valid
in the range of realistic experimental parameters. We have in 
this way presented an alternative analysis of the effects of
many-body interaction on EIT. As a prospect extension of the present
work, it would be
interesting to introduce a light mode to describe the probe laser in order
to explore the evolution of initially-coherent states of light and
the effect of particle interaction on photons.

The author thanks C. Adams and J. Otterbach for their comments and
C. Hadley for language advice. Financial support
from the project grants (R-144-000-276-112) and 
(R-710-000-016-271) is acknowledged.


\begin{thebibliography}{}


\bibitem{Marangos} M. Fleishchhauer, A. Imamoglu and J.P. Marangos, Rev. Mod. Phys. {\bf 77}, 633 (2005).

\bibitem{Renzoni} F. Renzoni, A. Lindner and E. Arimondo, Phys. Rev. A, {\bf 60}, 450 (1999).

\bibitem{Plotz} P. Pl\"otz, J. Madro\~nero and S. Wimberger, J. Phys. B: At. Mol. Opt. Phys. {\bf 43}, 081001 (2010).

\bibitem{Alexey} A.V. Gorshkov, J. Otterbach, M. Fleischhauer, T. Pohl and M. D. Lukin, arXiv:1103:3700.

\bibitem{Pritchard} J.D. Pritchard, A. Gauguet, K.J. Weatherill, M.P.A. Jones and C.S. Adams, Phys. Rev. Lett. {\bf 105}, 193603 (2010).

\bibitem{Schempp} H. Schempp, G. Gunter, C.S. Hofmann, C. Giese, S.D. Saliba, B.D. DePaola, T. Amthor, M. Weidem\"uller, S. Sevin\c cli and T. Pohl, Phys. Rev. Lett. {\bf 104}, 173602 (2010).

\bibitem{Wu} H.Z. Wu, Z.B. Yang and S.B. Zheng, Phys. Rev. A {\bf 82}, 034307 (2007).

\bibitem{Muller} M. M\"uller, I. Lesanovsky, H. Weimer, H.P. B\"uchler and P. Zoller, Phys. Rev. Lett. {\bf 102}, 170502 (2009).

\bibitem{Isenhower} L. Isenhower, E. Urban, X.L. Zhang, A.T. Gill, T. Henage, T.A. Johnson, T.G. Walker and M. Saffman. Phys. Rev. Lett. {\bf 104}, 010503 (2010).

\bibitem{efrain} E. Shahmoon, G. Kurizki, M. Fleischhauer and D. Petrosyan, Phys. Rev. A {\bf 83}, 033806 (2011).

\bibitem{inbal} I. Friedler, D. Petrosyan, M. Fleischhauer, G. Kurizki, Phys. Rev. A {\bf 72}, 043803 (2005).

\bibitem{Guerlin} C. Guerlin, E. Brion, T. Esslinger and K. M\o lmer, Phys. Rev. A, {\bf 82}, 053832 (2010).

\bibitem{yuan} H. Yuan, L.F. Wei, J.S. Huang, X.H. Wang and V. Vedral, arXiv:1101.2550.

\bibitem{Wilk} T. Wilk, A. Ga\"etan, C. Evellin, J. Wolters, Y. Miroshnychenko, P. Grangier and A. Browaeys, Phys. Rev. Lett. {\bf 104}, 010502 (2010).

\bibitem{Gillet} J. Gillet, G.S. Agarwal and T. Bastin, Phys. Rev. A {\bf 81}, 013837 (2010).

\bibitem{kwek} Y. Li, L. Aolita and L.C. Kwek, Phys. Rev. A {\bf 83}, 032313 (2011).

\bibitem{Moller} D. M\o ller, L.B. Madsen and K. M\o lmer, Phys. Rev. Lett. {\bf 100}, 170504 (2008).

\bibitem{Beterov} I.I. Beterov, D.B. Tretyakov, V.M. Entin, E.A. Yakshina, I.I. Ryabtsev, C. MacCormick and S. Bergamini, arXiv:1102.5223.

\bibitem{Saffman} M. Saffman, T.G. Walker and K. M\o lmer, Rev. Mod. Phys. {\bf 82}, 2313 (2010).

\bibitem{Otter} D. Petrosyan, J. Otterbach, and M. Fleischhauer, arXiv:1106.1360.

\bibitem{Henkel} S. Sevin\c{c}li, N. Henkel, C. Ates, and T. Pohl, arXiv:1106.2001.

\bibitem{Ates} C. Ates, S. Sevin\c{c}li and T. Pohl, Phys. Rev. A {\bf 83}, 041802 (2011).

\bibitem{Gong1} J. Gong and S. Rice, Phys. Rev. A, {\bf 69}, 063410 (2004).

\bibitem{Gong2} J. Gong and S. Rice, J. Chem. Phys., {\bf 120}, 5117 (2004).

\bibitem{Graefe} E.M. Graefe, H.J. Korsch and A.E. Niederle, Phys. Rev. Lett., {\bf 101}, 150408 (2008).

\bibitem{Brazil} V.S. Shchesnovich and D.S. Mogilevtsev, Phys. Rev. A, {\bf 82}, 043621 (2010).

\bibitem{Wieman} S.L. Cornish, N.R. Claussen, J.L. Roberts, E.A. Cornell and C.E. Wieman, Phys. Rev. Lett. {\bf 85}, 1795 (2000).

\bibitem{comment2} Here we abuse the standard notation and use the bra symbol to make reference to the left eigenvector of the NHH.

\bibitem{comment3} In order to facilitate the presentation of equations (\ref{eq:19}) and (\ref{eq:20}) we omit a normalisation factor of order $\Omega_p^2$. This is not the case when the same formulae are used in figure \ref{fig:4}.

\end{thebibliography}
\end{document}